\documentclass[12pt,preprint]{aastex}

\newcommand{\bdv}[1]{\mbox{\boldmath$#1$}}

\def\au{{\rm AU}} 

\def\masyr{{\rm mas}\,{\rm yr}^{-1}}
\def\rel{{\rm rel}}
\def\cen{{\rm cen}}
\def\geo{{\rm geo}}
\def\hel{{\rm hel}}
\def\mas{{\rm mas}}
\def\fwhm{{\rm FWHM}}
\def\e{{\rm E}}
\def\bpi{{\bdv\pi}}
\def\bmu{{\bdv\mu}}
\def\btheta{{\bdv\theta}}

\def\bbeta{{\bdv\beta}}

\def\muas{{\mu\rm as}}

\begin{document}
\title{Microlens Masses From Astrometry and Parallax in Space-Based Surveys:
 From Planets to Black Holes}
\author{
Andrew Gould\altaffilmark{1},
Jennifer C.\ Yee\altaffilmark{2,3}
}
\altaffiltext{1}{Department of Astronomy, Ohio State University,
140 W.\ 18th Ave., Columbus, OH 43210, USA}
\altaffiltext{2}{Harvard-Smithsonian Center for Astrophysics, 60 Garden St., 
Cambridge, MA 02138, USA}
\altaffiltext{3}{NASA Sagan Fellow}

\begin{abstract} 
We show that space-based microlensing experiments can recover
lens masses and distances for a large fraction of all events
(those with individual photometric errors $\la 0.01\,$mag)
using a combination of one-dimensional microlens parallaxes
and astrometric microlensing.  This will provide a powerful
probe of the mass distributions of planets, black holes,
and neutron stars, the distribution of planets as a function
of Galactic environment, and the velocity distributions of
black holes and neutron stars.  While systematics are in principle a 
significant concern, we show that it is possible to vet
against all systematics (known and unknown) using single-epoch
precursor observations with the {\it Hubble Space Telescope}
roughly 10 years before the space mission.

\end{abstract}


\section{{Introduction}
\label{sec:intro}}

At present, well under 1\% of microlensing events yield mass and distance
measurements.  It is estimated, for example, that 0.8\% of all events
(i.e., roughly 20 per year) are due to isolated black holes \citep{gould00},
but to date not a single one of these has been reliably identified as such.
Microlensing planet searches yield a dozen detections per year, a figure
likely to increase several fold over the next few years.  These planets
are distributed along the line of sight from near the Solar circle to
the Galactic bulge, and so potentially could tell us about planet
frequency as a function of Galactic environment.  In fact, distances
to these planets are mostly unknown, and those with measured distances
are highly biased toward being nearby.  Microlensing mass and distance
measurements could yield mass functions and velocity distributions
of black holes and neutron stars,
find detailed structures in the mass distribution of planets, identify
hosts by stellar type, and much more.

The main difficulty is that microlensing events routinely return
only one parameter that is sensitive to the mass and distance, namely
the Einstein timescale, 
\begin{equation}
t_\e = {\theta_\e\over \mu_\rel}; \quad
\theta_\e = \sqrt{\kappa M\pi_\rel} \simeq  0.3\,{\rm mas}
\sqrt{{M\over 0.5\,M_\odot}\,{\pi_\rel\over 20\muas}};
\quad
\kappa \equiv {4 G\over c^2\,\au}\simeq 8.1\,{{\rm mas}\over M_\odot}.
\label{eqn:tedef}
\end{equation}
Here $M$ is the lens mass, $\theta_\e$ is the Einstein radius,
$\pi_\rel$ is the lens-source relative parallax, and $\mu_\rel$
is the lens-source relative proper motion in the frame of the observer
(usually on Earth: $\mu_\rel=\mu_\geo$) at the peak of the event.

To disentangle the three physical parameters ($M,\pi_\rel,\mu_\rel$) 
that enter $t_\e$ clearly requires two additional observables.
For dark (or otherwise undetectable) lenses, these must be
the Einstein radius $\theta_\e$  and the {\it amplitude} of the
microlens parallax vector, $\pi_\e=|\bpi_\e|$.  This amplitude is
simply the trigonometric relative parallax scaled to the Einstein radius,
\begin{equation}
\pi_\e = {\pi_\rel\over\theta_\e}.
\label{eqn:piedef}
\end{equation}
The direction of $\bpi_\e$ is that of the lens-source relative motion,
$\bpi_\e/\pi_\e = \bmu_\rel/\mu_\rel$.

The usual path to measuring the lens mass and distance in the very few cases
that this has been done is to combine measurements of $\theta_\e$
and the {\it vector} $\bpi_\e$
\begin{equation}
(\theta_\e\,\oplus\,\bpi_\e) \Rightarrow M = {\theta_\e\over\kappa\pi_\e}
\quad \oplus \quad \pi_\rel=\theta_\e\pi_\e
\qquad {\rm (traditional)}.
\label{eqn:mdeval}
\end{equation}
Since the source parallax $\pi_s$ is usually well known
(typical uncertainty $\la 10\,\muas$), measuring
$\pi_\rel$ is equivalent to measuring the lens distance 
$D_l =\au/(\pi_\rel + \pi_s)$.

In this approach, $\theta_\e$ is only measurable for the subset of
events in which the source comes close to a caustic structure (typically
caused by a planetary or binary system).  The lightcurve then
yields $\rho=\theta_*/\theta_\e$, i.e., the ratio of the angular source radius
to the Einstein radius.  Since $\theta_*$ can be determined from the source
color and brightness \citep{yoo04}, one can then measure 
$\theta_\e=\theta_*/\rho$.  Relatively few events have such caustic anomalies.
Fortunately, this subset includes most planetary events.  However,
it includes extremely {\it few} black holes or neutron stars.

This latter fact is unfortunate because massive objects are the most
susceptible to microlens parallax measurements, which are the other
necessary ingredient.  The microlens parallax $\pi_\e=\pi_\rel/\theta_\e$
specifies the amplitude of lens-source relative displacement due
to reflex motion of the observer (typically on Earth) 
normalized to the Einstein radius.
If microlensing events lasted a year, this would induce very obvious annual
oscillations in the lightcurve.  But since most events are much shorter,
they contain only a fraction of an oscillation, which is usually
not detectable.  Even when detectable, it is usually only possible
to measure one component of the parallax vector, 
\begin{equation}
\pi_{\e,\parallel} \equiv \bpi_\e\cdot \hat{\bf n}_a,
\label{eqn:piepar}
\end{equation}
the component in the direction $\hat{\bf n}_a$ of the observer's
instantaneous acceleration (toward the projected position of the Sun)
at the peak of the event.  To the degree that $\hat{\bf n}_a$  is aligned
with $\bmu_\rel$, the event becomes asymmetric, rising
either faster or slower than it falls.  Since microlensing events
are intrinsically symmetric, asymmetric deviations are easily detected.
On the other hand, to the extent that $\bmu_\rel$ is perpendicular to
$\hat{\bf n}_a$, there is a symmetric distortion, which easily
masquerades as small changes in other symmetric parameters.  

Clearly, $\bpi_\e$ is easiest to measure when it is large and/or when
the timescale of the event is longer.  For example, massive objects
typically generate longer events, so there is more chance to measure
the full parallax effect. In practice,
mass and distance measurements are mainly made
for binary and planetary events (yielding $\theta_\e$) that happen
to be relatively close (so large $\pi_\rel$, implying large, more easily
measurable $\bpi_\e$).

Here we propose an alternate route to microlens mass measurements.
To do so, we introduce a new microlensing quantity, the vector Einstein
radius $\btheta_\e$
\begin{equation}
\btheta_\e\equiv \bmu_\rel t_\e;
\qquad \theta_{\e,\parallel} \equiv \btheta_\e\cdot \hat{\bf n}_a.
\label{eqn:btedef}
\end{equation}
We show that this new quantity is the observable in astrometric
microlensing and that it leads to a new path to microlens mass and distance
measurements:
\begin{equation}
(\btheta_\e\,\oplus\,\pi_{\e,\parallel})
\Rightarrow M = {\theta_{\e,\parallel}\over\kappa\pi_{\e,\parallel}}
\quad \oplus \quad \pi_\rel=\theta_\e^2{\pi_{\e,\parallel}\over\theta_{\e,\parallel}}
\qquad {\rm (new)}.
\label{eqn:mdeval2}
\end{equation}

In Section~\ref{sec:astrom}, we review astrometric microlensing and
show that what it actually measures is $\btheta_\e$.  
In Sections~\ref{sec:space} and \ref{sec:known}, we demonstrate that both 
$\btheta_\e$ and $\pi_{\e,\parallel}$ can be measured for a large fraction of events
detected in space-based microlensing surveys. In Section~\ref{sec:unknown},
we quantitatively evaluate problems posed by the known unknowns of this
approach, and in Section~\ref{sec:unknown2} we discuss the unknown unknowns.
The latter appear particularly intractable, but in  Section~\ref{sec:discuss}
we present an empirical method to control the 
unknowns, both known and unknown.

\section{{Review of Astrometric Microlensing}
\label{sec:astrom}}

Microlenses split source light into 2 or more images that are separated
on the sky by angles of order $\theta_\e$, which is typically $\la 1\,$mas.
Hence, these images are not typically resolved.  However, the centroid
of the combined image light is displaced from the source position by an
amount that scales directly as $\theta_\e$, and hence it is in principle
possible to measure $\theta_\e$ from a time series of astrometric measurements
\citep{my95,hnp95,walker95}.

For a point lens, there are two images with positions {\it relative to the lens}
$\Delta\btheta_\pm$ and magnifications $A_\pm$ given by \citep{einstein36}
\begin{equation}
\Delta\btheta_\pm = {u\pm\sqrt{u^2+4}\over 2}{\Delta\btheta\over u};
\qquad {\bf u}\equiv {\Delta\btheta\over \theta_\e};
\qquad \Delta\btheta \equiv \btheta_s - \btheta_l,
\label{eqn:thetapm}
\end{equation}
\begin{equation}
A_\pm = {A\pm 1\over 2},
\qquad A = {u^2+2\over u(u^2+4)^{1/2}},
\label{eqn:apm}
\end{equation}
where $\btheta_s$ and $\btheta_l$ are the astrometric positions of the source 
and lens.

Therefore, the displacement of the image centroid {\it from the source} is
\begin{equation}
\Delta\btheta_\cen = 
{A_+\Delta\btheta_+ + A_-\Delta\btheta_-\over A} - \Delta\btheta 
= {\Delta\btheta\over u^2 + 2} = {{\bf u}\over u^2 + 2}\theta_\e
\label{eqn:thetacen}
\end{equation}
If the lens-source relative motion is approximated as uniform
\begin{equation}
\Delta\btheta(t) = (t-t_0)\bmu_\rel + \bbeta\theta_\e
\label{eqn:traj}
\end{equation}
where $t_0$ is the time of closest approach and $\beta$ is the normalized
impact parameter $(\bbeta\perp\bmu_\rel)$, then $\Delta\btheta_\cen$
traces out an ellipse that is centered at 
$(\bbeta\theta_\e/2)/(\beta^2 + 2)$ and whose vector semi-major and 
semi-minor axes are
\begin{equation}
{\bf a} = {\btheta_\e\over 2(2 + \beta^2)^{1/2}};
\qquad
{\bf b} = {\bbeta\theta_\e\over 2(2 + \beta^2)}.
\label{eqn:alphabeta}
\end{equation}
Equation~(\ref{eqn:alphabeta}) verifies the claim made in 
Section~\ref{sec:intro} that astrometric microlensing measures $\btheta_\e$.
The axis ratio and eccentricity of this ellipse are
\begin{equation}
{a\over b} = \sqrt{{2\over \beta^2} + 1};
\qquad
e = \sqrt{2\over \beta^2 + 2}.
\label{eqn:axis}
\end{equation}

We note that the lens-source relative motion may not be rectilinear
for two reasons.  First,  either the source or lens may be accelerated
by a companion.  Second, even if both 
are intrinsically unaccelerated, the ellipse will be distorted by
parallax effects from the accelerated motion of the observer
\citep{boden98}.  We treat the first effect in Section~\ref{sec:unknown}.
The second effect is always be present at some level and so must be
included in formal fits.  However, it is generally quite small and
so can be ignored in the simplified treatment given here, which is
aimed at evaluating the viability of the method.

\section{{Application to Space-Based Surveys}
\label{sec:space}}

Space-based microlensing surveys have a number of advantages over ground-based,
but the most critical from the present perspective is the smaller 
point spread function (PSF).  This has three major implications.  
First, for sources above sky,
the astrometric precision is given by
\begin{equation}
\sigma_{\rm ast} =\sigma_{\rm phot}{\fwhm\over (\ln 256)^{1/2}}
\label{eqn:sigmaast}
\end{equation}
where $\sigma_{\rm phot}$ is the fractional photometric precision,
FWHM is the full width at half maximum of the PSF, and where we have
assumed a Gaussian PSF for definiteness (since the formula barely
changes for other plausible PSFs).  Since microlensing experimental
design is governed by requirements of photometric precision, this equation
automatically gives space a factor 5--10 advantage relative in astrometry
relative to the ground due to smaller PSF.  
Second, for the class of events relevant
for this approach, the sources are above sky from space, while 
most ground-based sources are below sky.  Finally, in high-resolution
space-based images, the source and lens are almost always isolated
from all unrelated stars, i.e., all stars other than possible companions
to these two stars.  This both facilitates precision astrometry and
immensely simplifies the analysis.

Nevertheless, despite the vastly improved astrometry from space, 
astrometric precision is still the limiting factor in mass measurements.
We will show below that the astrometric requirements imply that
$\sigma_{\rm phot}\la 0.01$.  This precision virtually
guarantees that $\pi_{\e,\parallel}$
will be well measured.  For example, Figure 3 from \citet{gould12}
shows the error ``ellipse'' for parallax in a simulated microlensing event with
a factor $\sim 2$ better precision than the typical one envisaged 
in this paper.  We put
``ellipse'' in quotation marks because it is difficult to make out
its width in the $\hat{\bf n}_a$ direction.  We will therefore assume
that $\pi_{\e,\parallel}$ is measured much better than $\btheta$.
Note, from Equation~(16) of \citet{gould12}, that $\pi_{\e,\parallel}$
precision deteriorates for shorter timescale events $\propto t_\e^{-2}$.
However, the shortest events are those 
for which astrometric effects are also extremely
difficult to measure.  Hence, while there may be some exceptions,
it is reasonable to assume that whenever $\btheta_\e$ can be measured,
$\pi_{\e,\parallel}$ will have been measured better.

\subsection{{Astrometric Microlensing Precision}
\label{sec:astroprecise}}

Similarly, but more so, for the events with high enough precision
to enable astrometric measurements, the basic event parameters 
$(t_0,\beta,t_\e)$ will be known with essentially infinite precision.
Therefore, the shape of the astrometric ellipse will likewise be known
with essentially infinite precision from the photometric data
(Equation (\ref{eqn:axis})).

Thus, six parameters are required to describe the image-centroid trajectory
in the neighborhood of the peak of the event:  the vector Einstein
radius $\btheta_\e$, the vector source proper motion $\bmu_s$,
and the vector source position at the peak of the event $\btheta_{s,0}$.
We will begin by approximating that $\bmu_s$ and  $\btheta_{s,0}$ as known
perfectly and test further below how well this holds
and under what conditions.

Then each astrometric measurement $k$ contributes to the determination of
the magnitude of $\theta_\e$ by
\begin{equation}
\sigma_k(\theta_\e) = \sigma_{{\rm ast},k}{u_k^2 + 2\over u_k},
\label{eqn:sigmak}
\end{equation}
where $\sigma_{{\rm ast},k}$ is the astrometric precision of the measurement
at separation $u_k$.
If the source is above sky, and assuming photon-limited statistics,
then $\sigma_{{\rm ast},k}=A_k^{-1/2}\sigma_{{\rm ast},0}$, where $\sigma_{{\rm ast},0}$ 
is the astrometric precision at baseline.  Then for a uniform series
of $N$ such measurements over a time interval $\Delta t=t_2-t_1$, the combined
precision is
\begin{equation}
\sigma(\theta_\e) = \sigma_{{\rm ast},0}\sqrt{\Delta \tau\over N}
\biggl[\int_{t_1}^{t_2} d\tau G[u(\tau)]\biggr]^{-1/2},
\label{eqn:sigmatot}
\end{equation}
where $\tau\equiv (t-t_0)/t_\e$, $\Delta\tau=\Delta t/t_\e$ and
\begin{equation}
G(u) \equiv \biggl({u\over u^2 + 2}\biggr)^2A(u)
= {u\over (u^2 + 2)\sqrt{u^2 + 4}}.
\label{eqn:gdef}
\end{equation}
Similarly, the error in $\phi$, the angular orientation of $\btheta_\e$, 
is given by $\sigma(\phi) = \sigma(\theta_\e)/\theta_\e$, so that
\begin{equation}
\sigma(\theta_{\e,\parallel}) = \sigma(\theta_{\e,\perp}) = \sigma(\theta_\e).
\label{eqn:sigmathetaepar}
\end{equation}
Explicitly,
\begin{equation}
\sigma(\theta_\e) = {\sigma_{{\rm ast},0}\over\sqrt{N\langle H \rangle}}
\rightarrow 28\,\muas
{\sigma_{{\rm phot},0}\over 0.01}\,
{\fwhm\over 175\,\mas}\,
\biggl({N \over 7000}\biggr)^{-1/2}
\biggl({\langle H \rangle \over 0.1}\biggr)^{-1/2}
\label{eqn:sigmatot2}
\end{equation}
where
\begin{equation}
\langle H \rangle \equiv 
{\tau_2 H(\tau_2;\beta) - \tau_1 H(\tau_1;\beta)\over \tau_2 - \tau_1};
\qquad
H(\tau;\beta) \equiv {1\over\tau}\int_0^\tau d\tau' G[u(\tau');\beta],
\label{eqn:hdef}
\end{equation}
and $H(-\tau;\beta) = H(\tau;\beta)$.  It is important to note that
$\sigma(\theta_\e)$ is larger than the combined astrometric precision
$N^{-1/2}\sigma_{\rm ast,0}$ by $\langle H \rangle^{-1/2}$
because the astrometric offset that is being measured is smaller than
$\theta_\e$ by this factor.  Hence, when comparing systematic to statistical
errors, it is the latter quantity 
$N^{-1/2}\sigma_{\rm ast,0}=\sigma(\theta_\e)\langle H \rangle^{-1/2}
\rightarrow 9\,\muas$ that
must be considered.

\begin{figure}
\plotone{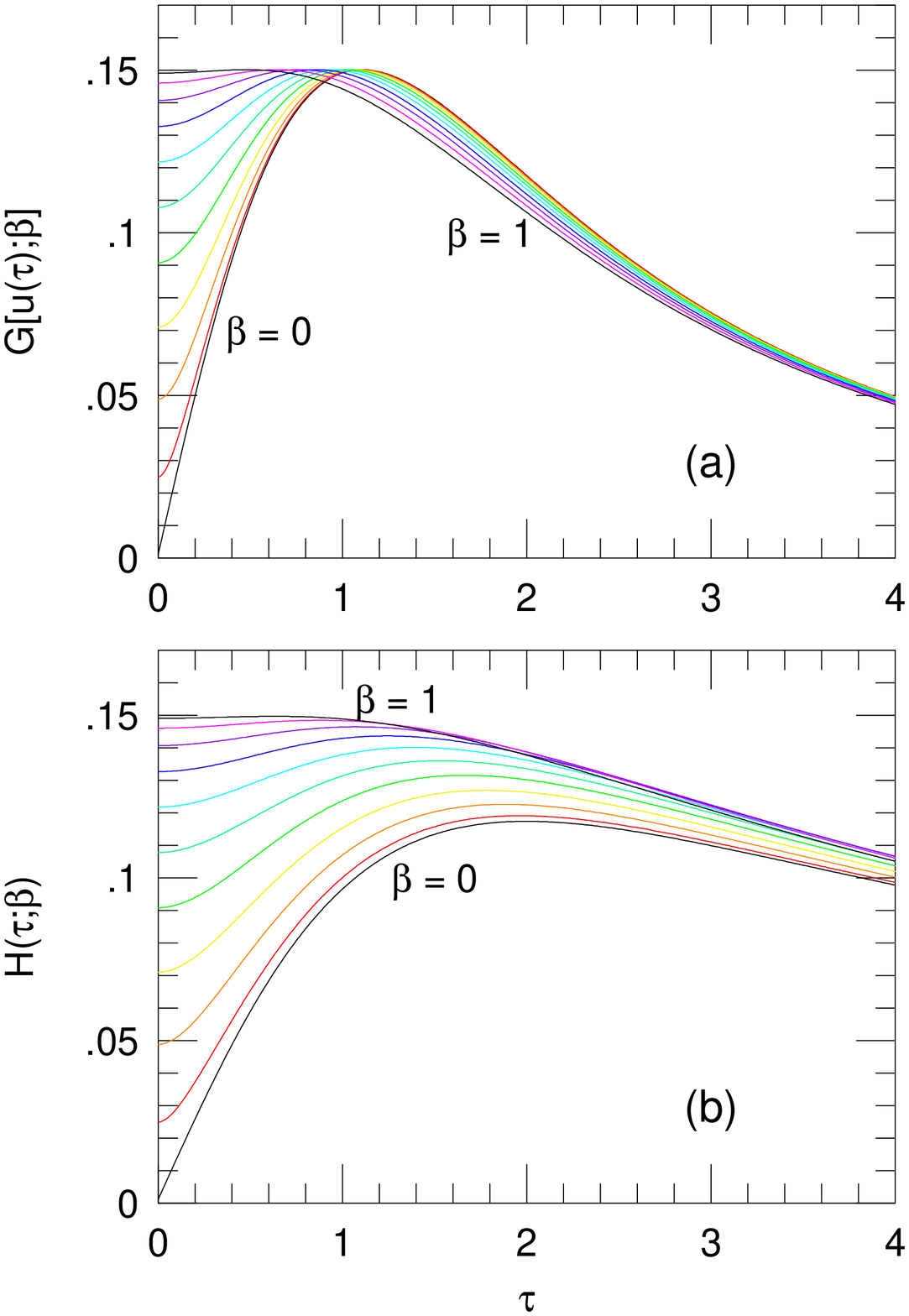}
\caption{
\label{fig:g}
Function $G$ and its integral $H$, whose expressions are given explicitly by
Equations~(\ref{eqn:gdef}) and (\ref{eqn:hdef}).  As shown
by Equation~(\ref{eqn:sigmatot2}) the error in $\btheta_\e$ 
scales as $\langle H \rangle^{-1/2}$, which according to 
Equation~(\ref{eqn:hdef}) is similar to 
$[H({\rm few},\beta)]^{-1/2}\sim 3$ for typical events.
}
\end{figure}

Figure~\ref{fig:g} shows the function $G[u(\tau);\beta]$ and
$H(\tau;\beta)$ for 11 values of $\beta$, $0\leq \beta \leq 1$.
It shows that for a very broad range of conditions, 
$\langle H\rangle\sim 0.1$.  Moreover, since this quantity enters
Equation~(\ref{eqn:sigmatot2}) only as the square root, the
astrometric precision is approximately independent of impact parameter
$\beta$, timescale $t_\e$, and time of closest approach $t_0$ over
this broad range (provided that the peak is within or near the
interval of observations).
That is, under the assumption of photon-limited
measurements, the indicated precision of $28\,\muas$ can typically
be achieved under the fiducial conditions, $N=7000$, $\fwhm=175\,\mas$,
and $\sigma_{{\rm phot},0}=1\%$.

These conditions have been scaled to one version of observations 
of the proposed {\it WFIRST} satellite, i.e., 94 observations per day
of each of 10 fields, over a continuous $\Delta t=72\,$day interval.
Simulations by M.\ Penny (2013, private communication), show that 
approximately half of the events with detectable Earth-mass planets would have
$\sigma_{{\rm phot},0} < 1\%$.  From Equation~(\ref{eqn:tedef}), typically
$\theta_\e\sim 0.3\,\mas$.  Such events would yield 10-sigma detections,
i.e., 10\% measurements of $\theta_\e$ under fiducial conditions.
Of course, what is required for mass determinations is a measurement
of $\theta_{\e,\parallel}$, and this quantity is smaller than $\theta_\e$
by the cosine of some random angle.  Nevertheless, this cosine will
be $>0.5$ in 67\% of all cases.  Hence, Equation~(\ref{eqn:sigmatot2})
implies that astrometric mass measurements are possible in a large
fraction of cases.

Before continuing, we note that for $\tau\ga 4$, all the curves in
Figure~\ref{fig:g}b tend toward $H(\tau;0)$, which can
be evaluated in closed form,
\begin{equation}
H(\tau,0) = 
{1\over\sqrt{2}\tau}
\ln {(\sqrt{\tau^2+4} -\sqrt{2})(1+\sqrt{2})\over \sqrt{\tau^2 +2}}
\rightarrow {0.62\over\tau}.
\label{eqn:hbeta0}
\end{equation}

\section{{Known Knowns}
\label{sec:known}}

\subsection{{Uncertainty in $\btheta_s$}
\label{sec:thetas}}

In Section~\ref{sec:space}, we evaluated the precision of the $\btheta_\e$
measurement under the assumption that the true position of the 
source $\btheta_s=\btheta_{s,0}+\bmu_s(t-t_0)$
was known with infinite precision, so that the measurement (and measurement
error) of its apparent position $\btheta_\cen$ directly gave the
offset between these, $\Delta\btheta_\cen = \btheta_\cen - \btheta_s$.
In fact, $\btheta_s$ must itself be determined from astrometric measurements,
which of course have similar individual precisions to those of the
$\btheta_\cen$ measurements.  Therefore, it is far from obvious that
the error in $\btheta_s$ can be ignored.

We address this issue in several phases.  We first assume that the lens
is dark and that
the source is isolated, i.e., that it has no massive companions
and so is moving in rectilinear relative motion.  Hence, the four
required parameters $\btheta_{s,0}$ and $\bmu_s$ can be determined
from a linear fit to data away from the event.  In practice, of course, 
one would fit all the data to a model with all the parameters.  However,
to determine the precision of these four parameters, we can idealize
the epochs in years other than the event as unaffected by the event.
We further idealize all the observations during each year as taking
place at the same time.  Assuming there are $n+1$ years of observations
(labeled $0,1,\ldots n$) with the event in the $j$th year, then
the inverse covariance matrix of each directional-component of the pair
($\btheta_{s,0},\bmu_s$) is given by (e.g., \citealt{gould03})
\begin{equation}
C^{-1} =
{N\over 6\sigma_{{\rm ast},0}^2}
\left(
\matrix{6n & 3(n+1)(n-2j){\rm yr}\cr
3(n+1)(n-2j){\rm yr} & (n+1)[n(2n+1)-6j(n-j)]{\rm yr}^2\cr}
\right),
\label{eqn:invcovmat}
\end{equation}
with e.g., $C_{11}^{1/2}=\sigma(\theta_{s,0,x})=\sigma(\theta_{s,0,y})$.
For simplicity, we restrict consideration to the least ($j=0$ or $j=n$) and
most $(j=n/2)$ favorable cases.  The first point to note is that
the error in the zero point dominates over the proper motion in either
case
\begin{equation}
{\sigma_\mu\over\sigma_\theta} = 
\sqrt{6\over (n+1)(2n+1)}{\rm yr}^{-1}\quad (j=0);
\qquad
{\sigma_\mu\over\sigma_\theta} = 
\sqrt{12\over (n+1)(n+2)}{\rm yr}^{-1}\quad (j=n/2).
\label{eqn:sigrats}
\end{equation}
That is, considering that for typical $n\sim 4$, these ratios are 
$<{\rm yr}^{-1}$, and that the source position needs to be known for
a time $\Delta t < 0.2\,{\rm yr}$, the proper motion errors
enter at an order of magnitude lower level than the positional errors.
{These positional errors can then be directly compared to the errors in
the mean offset estimated above, i.e., $\theta_\e\langle H\rangle^{1/2}$:
\begin{equation}
{\sigma(\theta_{s,0})\over\sigma(\theta_\e)\langle H\rangle^{1/2}} =
\sqrt{2(n+1)\over n(n -1)}
\quad (j=0);
\qquad
{\sigma(\theta_{s,0})\over\sigma(\theta_\e)\langle H\rangle^{1/2}} =
\sqrt{1\over n}.
\quad (j=n/2).
\label{eqn:errrat2}
\end{equation}
For example, for $n=4$, these
ratios are 90\% and 50\%, respectively.  Given that these errors enter
in quadrature, the second is sub-dominant but the first can be important.
Note, however, that for events whose peak is very roughly near the
center of a given year's observations, the role of the uncertainty
in the source position is greatly reduced by the symmetry in the
time evolution of the offset along the direction of motion, $\bmu_\rel$.
}

\subsection{{Luminous Lens}
\label{sec:luminous}}

If the lens is luminous, then the astrometric measurements at times
away from the event will yield the position and proper motion of
the combined lens and source light.  When these are projected back
to $t_0$, they will yield the lens-source centroid (as it would appear
if there were no astrometric lensing), with the same precision.
The division between lens and source light will be known very
precisely (under the assumption that neither has a companion) because
the microlensing fit precisely gives the source flux, and this
can be subtracted from the baseline flux to give the lens flux.

The main potential problem is that the amplitude of the astrometric
signal will be degraded because the source undergoes an astrometric
deviation but the blended light from the lens does not.  \citet{han99}
show that the combined light of the lens and source follows a
nearly-elliptical path, which is less eccentric than the
trajectory of the combined source images by themselves.  They conjecture
that this shape is truly an ellipse, although this has not to our
knowledge yet been proved.

However, if the lens is relatively faint, then this degradation is
minor and, as indicated above, with known functional form.  And if
the lens is bright, then its mass and distance can be estimated 
photometrically, so that the astrometric microlensing information is
relatively unimportant.

\section{{Known Unknowns}
\label{sec:unknown}}

It is known that stars in general frequently have binary companions, but 
it is unknown a priori whether this is the case for the particular 
lens and source in an event being studied.  Since such companions can
affect the interpretation of astrometric data, we classify them as
``known unknowns''.  

We consider first companions to the source.
The typical sources with sufficient S/N for astrometric measurements
are G dwarfs, and for these the companion is
typically so much fainter than the primary \citep{dm91}
that it can usually be considered ``dark''.  Accelerated source motion
due to such companions is usually called ``xallarap'' in the context
of photometric microlensing and we retain that terminology here.
The main concern is that this xallarap effect would go unrecognized
and would subtly corrupt the interpolation or extrapolation of the
source position
back to the time of the event.  For simplicity, we consider a 
face-on circular orbit. If the period is $P<5\,\rm yr$ (typical
duration of a space mission), then the amplitude of the astrometric signature
due to the companion will be simply
\begin{equation}
\Delta\theta_{\rm bin} = 30\,\muas\,
\biggl({P\over {\rm yr}}\biggr)^{2/3}
\biggl({M+m\over M_\odot}\biggr)^{1/3}
\biggl({m/(M+m)\over 1/4}\biggr)
\biggl({D_s\over {\rm 8\,kpc}}\biggr)^{-1}.
\label{eqn:orbit}
\end{equation}
Since the fiducial year-as-epoch astrometric error is about $9\,\muas$, this
implies that all such companions would be directly detectable unless they
had periods smaller than one year and/or had exceptionally low-mass.
At shorter periods $P\la 1\,$yr, the astrometric signature
would not, by itself, be secure.  However, such close companions typically
give rise to a photometric xallarap signature, even in ground-based data
(e.g., \citealt{poindexter05}), and the low-level astrometric signature
could be combined with photometric data to securely identify the xallarap
signature.

Let us then consider the opposite limit, in which the period is much
longer than the mission, so that the astrometric effect can be approximated
as uniform acceleration,
\begin{equation}
\alpha = 16\,\muas\,{\rm yr}^{-2}
\biggl({m\over (1/3)M_\odot}\biggr)
\biggl({a\over 10\,\au}\biggr)^{-2}
\biggl({D_s\over {\rm 8\,kpc}}\biggr)^{-1}
\sin\phi,
\label{eqn:uniform}
\end{equation}
where $a$ is the physical separation and $\phi$ is the angle relative
to the line of sight.  At, for example, $a\sim 10\,\au$, this signal
is too small to be reliably identified. However, if it is simply
ignored, this will lead to an incorrect estimate of $\btheta_{s,0}$ in
the direction of acceleration by
\begin{equation}
|\Delta\btheta_{s,0}| = {(n+1)(n+2)\over 12}\alpha\,{\rm yr}^2
\quad (j=0);
\qquad
|\Delta\btheta_{s,0}| = {(n+1)(n+2)\over 24}\alpha\,{\rm yr}^2
\quad (j=n/2).
\label{eqn:offset}
\end{equation}
To estimate the impact of these systematic errors, they should
be compared to the $\sigma(\theta_\e)\langle H \rangle^{1/2}\sim 10\,\muas$
statistical error.  Hence the systematics can be significant, but
only for about 0.5 dex of separations ($5\,\au\la a \la 15\,\au$), 
i.e., a few percent of all 
sources.  We return to the problem of evaluating the impact of
such systematic errors in Section~\ref{sec:discuss}.

Note that even if the source companion is extremely faint, it could
significantly displace the centroid of light if it were sufficiently
far away, thereby acting with a big 
``lever arm''.  For example, a companion that lay projected
at 500 AU and was 1/500 of the source brightness would displace the
centroid by $125\,\muas$ which is quite substantial compared to the 
quantities being measured.  However, this has no impact because
this displacement is essentially identical for all measurements and
so does not affect the {\it differential} astrometry, which is the
basis of the astrometric microlensing measurement.

We now turn to companions to the lens.  Just as with source companions,
the steepness of the mass-luminosity relation ensures that in most
cases the companion will have very different brightness from the lens.
However, in contrast to the source case, the lens companion is almost
as likely to be brighter than the lens as fainter.  This is because
microlensing event detection strongly selects on source brightness, but
it is essentially indifferent to lens brightness per se.  Therefore, the
only selection effect is the relative Einstein radii, which favors
the heavier of two well-separated companions by the square root of their
mass ratio.  The heavier of the two stars is usually also brighter, but
the (square root of the) mass ratio is typically tiny compared to the
luminosity ratio.
Even if the companion is brighter, this does not necessarily mean that
its light will be detectable.  However, it does mean that
in a substantial minority of all cases in which excess light is
detected, this excess will be due to a lens companion and not the
lens.  Nevertheless in the great majority of these cases, the lens
companion can be identified using the method of \citet{mb11293B}
wherein one measures the astrometric offset of the source at the
peak of the event (determined from difference imaging) relative to
the baseline source.  While the exact precision of this measurement
depends on the peak magnification and event timescale, typically
the precision will be of order $20\,\muas$ for data of sufficient
quality to do astrometric microlensing.  Even if the companion is
just 10\% of the source brightness, this implies that it will be
detectable if it lies more than a few AU from the lens.  In the remaining
cases, the companion will lie inside (or at least very near) the Einstein
radius of the lens and therefore will give rise to pronounced astrometric
effects that will be of interest to measure \citep{han99b,han01,gouldhan}.
It will also very likely give rise to photometric (binary lensing)
effects.

\section{{Unknown Unknowns}
\label{sec:unknown2}}

Even very small systematic errors could in principle radically undermine 
the mass determinations derived from astrometric microlensing.
The precision of the masses relies on ``root-n'' scaling
of the statistical error bars in which $n\sim 10^8$ photons
(i.e. $N\sim 7000\,{\rm yr}^{-1}$ observations
each with ${\rm (S/N)^2}\sim 10^4$). However, systematic errors do
not scale this way.
As an example, if the mean of the astrometric reference frame were 
displaced from the source by just $1^{\prime\prime}$, the differential
aberration of light would be ${\cal O}(10\,\muas)$.  Of course, one
does not expect differential aberration to be ignored in the analysis,
but the point is that other very tiny effects in the optics, the 
detector, etc., could impact the result.  Happily, in the next section, 
we discuss ways to control for systematic errors of the most general sort.

\section{{Systematic Vetting of Unknowns (Known and Unknown)}
\label{sec:discuss}}

Given that these mass measurements will be both near the detection limit
and dependent on near-perfect scaling of root-n errors, the only method
for assessing their viability is to make independent proper motion 
measurements of a substantial subsample.  At first sight this appears to
be a daunting task because the whole driver of the approach we have outlined is
to measure the masses of objects that are otherwise unmeasurable.
However, as mentioned in the Introduction, there is already a widely used 
technique for measuring the scalar proper motion $\mu_\geo$ for events with
caustic crossings, which includes most planetary and binary events.
These are also among the events of greatest direct interest.  

Of course, as we have emphasized, the mass determinations discussed here
require measurement of the {\it vector} proper motion $\bmu_\geo$, not
just its amplitude.  However, if the amplitudes were correctly
estimated from astrometric microlensing (within errors) it would
be strong evidence that the directions were measured correctly as
well, since the amplitude and direction are approximately independent
and derive from the same quality of data.
This provides one check on the astrometric proper motion measurements.

Nevertheless, from the discussion in Section~\ref{sec:unknown}, it is
guaranteed that systematics will corrupt at least some measurements and,
as emphasized in Section~\ref{sec:unknown2} the level of this problem
is very difficult to assess in advance.  Therefore, it would be valuable
to have independent measurements of the vector proper motions (and for
a much larger subsample than the scalar proper motions discussed above)
in order to be able to systematically study
the conditions under which the astrometric measurements are significantly
corrupted and, hopefully, to identify the source of these problems and
find means to ameliorate them.

This can be accomplished via a high-resolution survey of future
microlensing fields using the {\it Hubble Space Telescope (HST)}.  In a 
subset of cases that we identify below, the future lens and source
will be separately resolved in such images.  Since the precise time of their
near-perfect future alignment will be determined by the microlensing
event, it will be possible to measure their heliocentric proper motion
$\bmu_\hel$.  This is related to their geocentric motion by
\begin{equation}
\bmu_\hel = \bmu_\geo + {{\bf v}_{\oplus,\perp}\over \au}\pi_{\rm rel}
\label{eqn:helgeo}
\end{equation}
where ${\bf v}_{\oplus,\perp}$ is the motion of Earth relative to the Sun
projected on the plane of the sky at the peak of the event, which is
known extremely well. Note that $\pi_\rel$ can be estimated either
by photometric parallax of the separately resolved lens or from the
overall microlensing solution: $\pi_\rel = \theta_\e\pi_\e$.

There are several challenges to making such measurements.  First, of
course, to be separately resolved the lens must be luminous.  This
already precludes using this test on dark lenses.  However, as we have
argued above, if the lens is a factor several fainter than the source,
the astrometric measurements are only mildly affected and in a well-understood
way.  Hence the approach could plausibly be applied to lenses that are a
factor $\sim 3$ to $\sim 20$ fainter than the source.

More challenging, the lens and source must be separately resolved at the time
of the earlier observations.  This requirement will vary as a function of
system parameters, but for definiteness, we adopt the FWHM as the minimum
separation.  This immediately leads to the proper motion requirement
\begin{equation}
\mu_\hel > \mu_{\rm min} = {\fwhm\over \Delta t} = 8.4\,\masyr
{\lambda\over 0.8\,\mu{\rm m}}\biggl({\Delta t\over 10\,{\rm yr}}\biggr)^{-1},
\label{eqn:hst}
\end{equation}
where $\lambda$ is the central wavelength of the observing passband and
$\Delta t$ is the elapsed time between the {\it HST} observations and
the microlensing event.  Note that this minimum is substantially
higher than the typical $\mu_\geo$ measured for microlensing events
\citep{henderson14}.

While at first sight one might think that these advance {\it HST} 
observations should be done in a similar passband to the microlensing 
observations, Equation~(\ref{eqn:hst}) makes clear that there is a 
huge premium on going to shorter wavelengths.  In particular, if
we consider bulge lenses and assume isotropic bulge-star
proper motion dispersions
$\sigma_\mu \ll \mu_{\rm min}$, then the fraction of events satisfying
Equation~(\ref{eqn:hst}) is just
\begin{equation}
f = {2\over\sqrt{\pi}} x\exp(-x^2)
\qquad \biggl(x\equiv {\mu_{\rm min}\over 2\sigma_\mu}\biggr).
\label{eqn:fval}
\end{equation}
If we then adopt $\sigma_\mu=3\,\masyr$, $\Delta t=10\,$yr, 
and $\lambda=\lambda_I=0.8\,\mu$m, then $f= 22\%$.  
But for the same assumptions and $\lambda=\lambda_H=1.65\,\mu$m, 
$f< 0.1\%$.  Fortunately, however, initial work by M.\ Penny (2013,
private communication) shows that the best {\it WFIRST} fields 
have significant (roughly 50\%) overlap with current OGLE fields where the
extinction is low enough that many microlensing events are currently
being found in $I$-band.  On the other hand, the several additional
magnitudes of extinction in $V$ band preclude going to substantially
shorter wavelengths.  Therefore, early observations of these overlapping
OGLE and  {\it WFIRST} fields should be carried out in $I$ band.

\acknowledgments

Work by AG was supported by NSF grant AST 1103471 
and NASA grant NNX12AB99G.  Work by JCY was performed in part under 
contract with the California Institute of Technology (Caltech) funded 
by NASA through the Sagan Fellowship Program.

\end{document}